# Disciplinary Variations in Altmetric Coverage of Scholarly Articles

Sumit Kumar Banshal[1], Vivek Kumar Singh[2], Pranab K. Muhuri[3] and Philipp Mayr[4]

[1] sumitbanshal06@gmail.com
Department of Computer Science, South Asian University, New Delhi (India)

[2] vivek@bhu.ac.in
Department of Computer Science, Banaras Hindu University, Varanasi (India)

[3] pranabmuhuri@gmail.com
Department of Computer Science, South Asian University, New Delhi (India)

[4] philipp.mayr-schlegel@gesis.org
GESIS Leibniz Institute for Social Sciences, Cologne (Germany)

## Abstract
The popular social media platforms are now making it possible for scholarly articles to be shared rapidly in different forms, which in turn can significantly improve the visibility and reach of articles. Many authors are now utilizing the social media platforms to disseminate their scholarly articles (often as pre- or post- prints) beyond the paywalls of journals. It is however not very well established if the level of social media coverage and attention of scholarly articles is same across all research disciplines or there exist discipline-wise variations. This paper aims to explore the disciplinary variations in coverage and altmetric attention by analyzing a significantly large amount of data from Web of Science and Altmetric.com. Results obtained show interesting patterns. Medical Sciences and Biology are found to account for more than 50% of all instances in Altmetrics. In terms of coverage, disciplines like Biology, Medical Science and Multidisciplinary Sciences have more than 60% of their articles covered in Altmetrics, whereas disciplines like Engineering, Mathematics and Material Science have less than 25% of their articles covered in Altmetrics. The coverage percentages further vary across different altmetric platforms, with Twitter and Mendeley having much higher overall coverage than Facebook and News. Disciplinary variations in coverage are also found in different altmetric platforms, with variations as large as 7.5% for Engineering discipline to 55.7% for Multidisciplinary in Twitter. The paper also looks into the possible role of source of publication in altmetric coverage level of articles. Interestingly, some journals are found to have a higher altmetric coverage in comparison to the average altmetric coverage level of that discipline.

## Introduction
The rapid growth of the Internet and social media has not only transformed the businesses, organizations and society, but has also changed the entire process of scholarly information processing, including article storage, access and dissemination. Not only research articles are being stored in and accessed from online digital libraries, but they are also disseminated through different social media platforms. Scholarly articles are now disseminated and shared on different social media platforms such as ResearchGate, Twitter, Facebook etc. There are some other popular platforms dedicated mainly to dissemination and sharing of academic texts, such as Academia and Mendeley. These academic networks provide wide-range of facilities which can be useful for academics (Gruzd & Goertzen, 2013). Several studies (such as by Williams & Woodacre, 2016) have found these types of academic social networks very informative and relevant for quantitative characterization in research assessments. This social media phenomenon of scholarly articles has become so popular that now a new range of metric have been designed, called alternative metric, to measure the interaction of social media with scholarly information processing (Priem, 2014; Priem & Hemminger, 2010).

Altmetrics is now an interesting area of research where researchers try to analyze the social media coverage and consumption of scholarly articles; and sometimes Altmetric values are even able to predict future citations. However, most of the attention of research in the area has so far been concentrated on measuring correlations and interactions among social media transactions and citation behaviour of scholarly articles. Relatively lesser attention has been paid on measuring disciplinary variations in social media coverage and usage of scholarly articles. This paper tries to address the issue through a comprehensive study involving large amount of data collected from Web of Science and corresponding values from Altmetrics. The main objective is to find out if there exist discipline-wise variations in social media coverage and consumption patterns of the scholarly articles. The data for different platforms (namely Twitter, Facebook, News and Mendeley) is analysed computationally for the purpose. Statistics for some highly visible journals in Social Media and journals with high impact factor are also analysed to understand the role of source of publication and disciplinary variations.

**Related Work**

There has been some attention of researchers on understanding and analyzing the relationship of social media and scholarly information systems. Some of these studies (Priem, 2014; Haustein et al., 2014; Thelwall & Kousha, 2015; Sugimoto et al., 2017) tried to understand and demonstrate if social media platforms can be used (or not) as a tool to attract more attention towards a published work. Few others (Shema, Bar-Ilan & Thelwall, 2014; Thelwall, 2016; Peters et al., 2016) tried to see if Altmetrics could correlate with citations, with few (Costas, Zahedi & Wouters, 2015a) going to the extent to see if it can complement citations or not. There have also been studies that tried to predict early citations from different platforms of social media, such as Mendeley (Thelwall, 2018), ResearchGate & Google Scholar (Thelwall & Kousha, 2017 a), altmetric.com (Thelwall & Nevill, 2018), and CiteULike bookmarks (Sotudeh, Mazarei & Mirzabeigi, 2015) etc. Country specific Altmetric studies has also been done, such as for India (Banshal et al., 2018) and China (Wang et al., 2016) etc.

Discipline-specific studies of understanding Altmetric coverage and impact have been done by some researchers. For example, a study by Bar-Ilan (2014) mapped astrophysics research output with Mendeley readership behavior using Scopus and Arxiv. Another study (Sotudeh, Mazarei & Mirzabeigi, 2015) analyzed the correlation of research impact and CitedULike bookmarks in Library & Information Science discipline. Few other such efforts are analyzing the relationship between traditional and alternative matrices in psychology literature for the period of 2010-2012 (Vogl, Scherndl & Ku, 2018); online media presence of Swedish articles in humanities in the year 2012 (Hammarfelt, 2014); and evaluation of the impact of Altmetrics in social sciences and humanities research published by Taiwan based researchers (Chen et al., 2015). In a recent work (Htoo & Na, 2017) worked towards alternative metrics across various disciplines of Social Sciences and visualized the significance of ten selected indicators on nine disciplines of social science. However, there has been relatively less attention on understanding disciplinary variations in altmetric coverage of scholarly articles.

The only past works found on Altmetrics with focus on disciplinary analysis are as follows: Holmberg & Thelwall (2014) conducted a study on data from Twitter of ten selected disciplines to map their coverage and frequencies in twitter. Authors here selected ten different disciplines based on their publication size and pattern variations to represent variations in publishing scholarly communication. Similar to this work, authors selected ten disciplines of social sciences and humanities from web of science subject areas to correlate with Mendeley readership (Mohammadi & Thelwall, 2014). However, analyzing overall disciplinary variations in coverage was not the main objective of the paper. Another work (Zahedi, Costas & Wouters, 2014) used a multi-disciplinary approach on different online and social networks to assess coverage and distribution of randomly selected 20,000 articles published between 2005 and 2011. This approach also outlined alternative metrics into seven different broader areas of research. These seven broader areas are classified based on high level classification which classified the research areas as ‘Natural Sciences’, ‘Engineering Sciences’ etc. In another related work (Costas, Zahedi & Wouters, 2015b), authors tried to understand the thematic orientation of publications mentioned on social media. However, this paper only uses altmetric data and tries to understand the distribution of the data into various high-level disciplines. It does not measure coverage levels of different disciplines in altmetric. Another work (Ortega, 2015) used thousands of Spanish researchers’ profiles to explore the disciplinary behavioral patterns in three online media, namely ResearchGate, Academia & Mendeley. Furthermore, in this analysis the scholarly articles are classified into eight broader areas to visualize the presence and coverage of same discipline’s researchers across different platforms. In one relatively recent work (Thelwall & Kousha, 2017b) scholarly communications shared in ResearchGate is being classified into 27 Scopus categories of subject area where the Scopus subject areas are used as it is defined to classify the articles and understand the disciplinary variations.

The present work has focused objective of analyzing coverage levels of articles from different disciplines in Altmetrics. It uses a large amount of data from Web of Science (about 1.4 million records to be precise) and their corresponding entries in Social media platforms. The objective is to understand whether research articles from all disciplines get equal coverage in social media platforms or not. Data is categorized into 14 different well-identified broader research areas/ disciplines and variations in altmetric coverage across these disciplines are identified. Unlike, previous studies the present work mainly tries to analyze altmetric coverage levels of different disciplines and not the disciplinary distribution of altmetric data, pursued by many of the previous studies. Further, a journal-based analysis is also done to understand the disciplinary variation and its impact on altmetric coverage.

## Data

The data is obtained from two sources: Web of Science and Altmetric.com. First of all the publication records for the year 2016 are downloaded from Web of Science. The data download process is performed during 1-10th December, 2018. A total number of 2,528,868 publication records are found for the year 2016. This data is then scanned for DOI entries and those records that do not have DOI are removed. This process reduces the data to 1,460,124 records. The second step in data collection involved collecting altmetric data for the 1,460,124 publication records of Web of Science with DOIs. For this purpose, the popular portal altmetric.com was accessed. In altmetric.com, 18 different types of mentions and stats are provided. These comprise of different social network mentions and reads. Out of the 1,460,124 records, a total of 681,274 publication records are found indexed in altmetric.com. Out of these 650,009 records are found with at least one kind of statistics. This corresponds roughly to 45% of data collected from Web of Science having DOI. Though altmetric.com captures statistics from various social platforms; platforms like Twitter, Facebook, News and Mendeley are found to be more popular. We have, therefore, used the altmetric data for these four platforms. The analysis also involves Impact Factor data for different journals, which is collected from Web of Science Reports.

## Disciplinary Tagging

To understand disciplinary variations in altmetric coverage, it is necessary to tag each publication record with at least one specific discipline. For this task, each publication record in the dataset is classified into one of the 14 broad research disciplines, as proposed in an earlier work (Rupika et al., 2016). This tagging is done by using Web of Science Category (WC) field information. One record can be tagged with multiple disciplines of research based on its WC entries. These 14 broader disciplines are as follows: Agriculture (AGR), Art & Humanities (AH), Biology (BIO), Chemistry (CHEM), Engineering (ENG), Environment Science (ENV), Geology (GEO), Information Sciences (INF), Material Science (MAR), Mathematics (MAT), Medical Science (MED), Multidisciplinary (MUL), Physics (PHY) and Social Science (SS). Thus the 255-category division of articles in Web of Science is reduced to these 14 broader disciplines and each publication record is tagged with one (or more) broad disciplines. All further analysis on disciplinary variations in altmetric coverage are done across these 14 broad disciplines. The articles are grouped discipline-wise and analytical results are obtained accordingly.

### Disciplinary Distribution and Coverage

The first point of analysis was to find out disciplinary distribution of articles in Web of Science & Almetrics and to see if disciplines are distributed in same proportions in Web of Science & Altmetrics. **Figure 1** presents the discipline-wise distribution of research output in Web of Science and altmetric.com. We can observe that some disciplines with higher proportionate distribution in Web of Science have relatively lesser proportion of presence in altmetric.com. In contrast, the Medical Science (MED) discipline accounts for about 30.2% proportion of output in Web of Science whereas in altmetric.com, it accounts for more than 41% of articles covered. Thus, this discipline is over covered in Altmetrics. For many other disciplines, proportionate contribution in Web of Science and altmetrics.com are different. For example, in Web of Science, PHY has the second most published output with contribution of 13.8% whereas its proportionate contribution in altmetrics.com is 7.9%. In Altmetrics, Social Sciences (SS) has the second highest proportionate contribution with a share of 15.4% followed by Biology with a share of 11.9%. Thus, MED and BIO disciplines are more visible in altmetric coverage, being covered in proportion higher than their proportion of published articles indexed in Web of Science. Disciplines like PHY, MAR, MAT, INF and ENG are proportionately less covered in Altmetrics. These results indicate that difference in altmetric coverage proportion of different disciplines are likely.

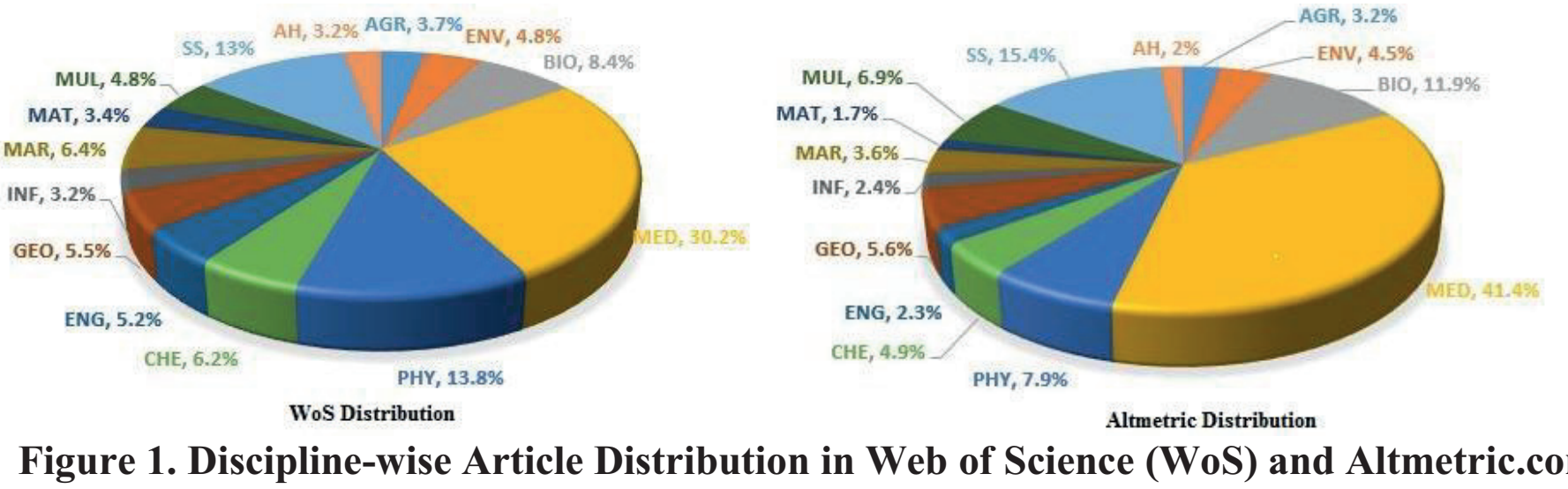

**Figure 1. Discipline-wise Article Distribution in Web of Science (WoS) and Altmetric.com**

The second point of analysis was to look at each discipline and find out its coverage level in Altmetrics. For this purpose, the Web of Science article counts for different disciplines is taken and altmetric.com is searched to see if they are covered in Altmetrics. This is done through an article-wise lookup in altmetric.com, for each article in Web of Science. **Table 1** shows the counts of articles of different disciplines that are indexed in Web of Science, number of articles found in altmetrics.com, and the altmetric coverage percentage for each of the 14 disciplines. It is observed that there is a significant difference in altmetric coverage percentages. For example, MUL, MED and BIO disciplines have a coverage percentage above 60%, which shows that out of all publications from these disciplines in Web of Science, more than 60% are found covered in Altmetrics. Articles from SS discipline have a coverage percentage of more than 50%. Interestingly, disciplines like ENG, MAT and MAR have less than 25% of their articles covered in Altmetrics. There is, therefore, a clear disciplinary variation in altmetric coverage of articles.

**Table 1. Discipline-wise data for altmetric coverage of articles indexed in Web of Science (WoS)**

| *Discipline* | *Articles in WoS* | *Altmetric Presence* | *Coverage Percentage* |
|---|---|---|---|
| AGR | 53749 | 21068 | 39.2 |
| AH | 47186 | 12871 | 27.3 |
| BIO | 123180 | 77259 | 62.7 |
| CHE | 90959 | 31670 | 34.8 |
| ENG | 75834 | 14737 | 19.4 |
| ENV | 69709 | 29194 | 41.9 |
| GEO | 80477 | 36420 | 45.3 |
| INF | 46438 | 15568 | 33.5 |
| MAR | 94117 | 23571 | 25 |
| MAT | 49385 | 10865 | 22 |
| MED | **441032** | **268830** | 61 |
| MUL | 69445 | 44778 | **64.5** |
| PHY | 201373 | 51569 | 25.6 |
| SS | 189835 | 100029 | 52.7 |

The third point of analysis was to find out if the discipline-wise coverage patterns are same across different altmetric platforms or if they vary significantly. For this purpose, coverage patterns across four different altmetric platforms, namely Twitter, Facebook, News and Mendeley are identified. **Table 2** shows the data for coverage of articles indexed in Web of Science in different altmetric platforms, corresponding to different disciplines. It is observed that, Mendeley and Twitter have in general higher coverage percentage for most of the disciplines as compared to News and Facebook. Thus, Mendeley and Twitter appear to be more popular altmetric platforms. It can also be observed from the table that there are noticeable disciplinary differences in altmetric coverage of articles. For example, articles from MUL discipline have highest presence in both Twitter & Mendeley with 55.7% and 63.6% coverage followed by BIO with 54.6% & 62.1% coverage. But articles from disciplines like ENG and INF have less coverage in Twitter (7.5% and 9.3%, respectively) and Mendeley (18.6% and 30.5%, respectively). In case of Facebook and News, coverage levels are low, with highest coverage being for MUL discipline of 17.8% followed by MED discipline of 13.7%. In News platform, coverage levels are further low with highest being 13% for MUL discipline followed by 7.5% for MED discipline. Interestingly, articles from ENG discipline have low coverage (7.5% in Twitter; 1.4% in Facebook; 18.6% in Mendeley and 0.5% in News) across all the platforms.

In terms of variations across disciplines, Twitter has the largest variation in coverage ranging from low of 7.5% for ENG to 55.7% for MUL. The variation range is in Mendeley is from 18.6% for ENG to 63.6% for MUL discipline, almost similar as Twitter. Facebook has variation in coverage percentage ranging from 1.3% for INF to 17.8% for MUL discipline. Thus, it is clearly observed that there exist disciplinary variations in altmetric coverage of articles, which varies further across different altmetric platforms. It may also be interesting to see if these variations can be attributed mainly to disciplines or if there are other factors such as the source of publication (journal), which play an important role.

**Table 2. Discipline-wise Coverage across Different Platforms of Articles indexed in Web of Science (WoS)**

| *Discipline* | *Articles in WoS* | *Twitter* | | *Facebook* | | *News Mention* | | *Mendeley* | |
|---|---|---|---|---|---|---|---|---|---|
| | | *#of Articles* | *Coverage Percentage* | *#of Articles* | *Coverage Percentage* | *#of Articles* | *Coverage Percentage* | *#of Articles* | *Coverage Percentage* |
| AGR | 53,749 | 16,132 | 30 | 4,406 | 8.2 | 1,468 | 2.7 | 20,784 | 38.7 |
| AH | 47,186 | 8,690 | 18.4 | 2,025 | 4.3 | 350 | 0.7 | 10,763 | 22.8 |
| BIO | 123,180 | 67,281 | 54.6 | 16,850 | 13.7 | 9,006 | 7.3 | 76,480 | 62.1 |
| CHE | 90,959 | 24,733 | 27.2 | 4,673 | 5.1 | 2,332 | 2.6 | 31,331 | 34.4 |
| ENG | 75,834 | 5,663 | 7.5 | 1,067 | 1.4 | 355 | 0.5 | 14,128 | 18.6 |
| ENV | 69,709 | 22,196 | 31.8 | 4,722 | 6.8 | 2,219 | 3.2 | 28,961 | 41.5 |
| GEO | 80,477 | 26,873 | 33.4 | 5,445 | 6.8 | 3,599 | 4.5 | 35,902 | 44.6 |
| INF | 46,438 | 4,330 | 9.3 | 583 | 1.3 | 373 | 0.8 | 14,151 | 30.5 |
| MAR | 94,117 | 15,096 | 16 | 2,508 | 2.7 | 1,674 | 1.8 | 23,280 | 24.7 |
| MAT | 49,385 | 5,773 | 11.7 | 792 | 1.6 | 618 | 1.3 | 9,777 | 19.8 |
| MED | **441,032** | **224,132** | 50.8 | **70,401** | 16 | 33,021 | 7.5 | **264,405** | 60 |
| MUL | 69,445 | 38,675 | **55.7** | 12,371 | **17.8** | 9,021 | **13** | 44,194 | **63.6** |
| PHY | 201,373 | 33,571 | 16.7 | 5,973 | 3 | 3,908 | 1.9 | 50,031 | 24.8 |
| SS | 189,835 | 78,799 | 41.5 | 24,557 | 12.9 | 9,258 | 4.9 | 96,180 | 50.7 |

## Analysing Disciplinary Variations by Journals

It is quite clear from the discussion in previous section that there are disciplinary variations in altmetric coverage of scholarly articles. An important and relevant question worth exploring here would be to find out if the source of publication (i.e. journal) has any role in higher altmetric coverage of articles. In order to explore this question, a part of data was taken out and analysed. This data comprised of top 100 journals (ranked by Web of Science article count) with the condition that they should have at least 500 articles covered in Altmetrics. These journals are then tagged with a primary discipline, based on data available either on their homepage or Wikipedia. Thus, each journal is categorized into one of the 14 broad disciplines. **Table 3** presents the data for these journals. It can be observed that MED discipline accounts for 35 out of these 100 journals followed by BIO with 24 journals and CHEM with 20 journals. These three disciplines taken together account for about 80% of the top 100 journals. In terms of coverage, MUL discipline has highest number of papers covered in all the four altmetric platforms, though it has only 6 out of 100 journals. In terms of coverage percentage, ENG discipline has the highest coverage in Twitter (86.2%) and Mendeley (85.6%) followed by GEO discipline. Disciplines like MED and BIO have somewhat lesser, but still a significant coverage of articles in Altmetrics. For example, MED has coverage percentage of 59.3% in Twitter and 65% in Mendeley for its articles in the selected sample. Similarly, BIO discipline has coverage percentage 62.2% in Twitter and 69.1% in Mendeley. However, the lesser overall covered disciplines like INF, MAT, MAR are found better covered in this sample. Thus, it is very difficult to conclusively say that publication in a particular journal gives an article a higher chance of altmetric coverage. The disciplinary variations are, however, still seen.

**Table 3. Disciplinary Distribution of 100 Most Productive Journals (ranked by WoS count) across Platforms**

| *Discipline* | *#of Journals* | *# Articles in WoS* | *Twitter* | | *Facebook* | | *News Mention* | | *Mendeley* | |
|---|---|---|---|---|---|---|---|---|---|---|
| | | | *#of Articles* | *Coverage %* | *#of Articles* | *Coverage %* | *#of Articles* | *Coverage %* | *#of Articles* | *Coverage %* |
| MED | **35** | **42,988** | 25,476 | 59.3 | 7,922 | 18.4 | 4,724 | 11 | 27,955 | 65 |
| BIO | 24 | 30,596 | 19,027 | 62.2 | 4,380 | 14.3 | 3,734 | 12.2 | 21,127 | 69.1 |
| CHE | 20 | 46,407 | 17,522 | 37.8 | 2,986 | 6.4 | 1,828 | 3.9 | 21,437 | 46.2 |
| PHY | 18 | 39,474 | 13,941 | 35.3 | 2,775 | 7 | 1,483 | 3.8 | 17,748 | 45 |
| MUL | 6 | 46,502 | **29,894** | 64.3 | **8,360** | 18 | **8,205** | **17.6** | **33,738** | 72.6 |
| ENV | 4 | 6,306 | 2,363 | 37.5 | 287 | 4.6 | 284 | 4.5 | 3,000 | 47.6 |
| GEO | 3 | 2,897 | 2,279 | 78.7 | 271 | 9.4 | 221 | 7.6 | 2,401 | 82.9 |
| MAR | 3 | 4,149 | 1,576 | 38 | 202 | 4.9 | 373 | 9 | 2,125 | 51.2 |
| AGR | 2 | 1,694 | 1,240 | 73.2 | 328 | 19.4 | 157 | 9.3 | 1,360 | 80.3 |
| SS | 2 | 1,270 | 953 | 75 | 474 | **37.3** | 67 | 5.3 | 1,040 | 81.9 |
| ENG | 1 | 1,149 | 990 | **86.2** | 8 | 0.7 | 11 | 1 | 983 | **85.6** |
| INF | 1 | 793 | 482 | 60.8 | 67 | 8.4 | 76 | 9.6 | 524 | 66.1 |
| MAT | 0 | 0 | 0 | 0 | 0 | 0 | 0 | 0 | 0 | 0 |
| AH | 0 | 0 | 0 | 0 | 0 | 0 | 0 | 0 | 0 | 0 |

For a more detailed analysis, the journals and their data are arranged in a different ranking order. **Table 4** shows the data for top 50 journals, ranked both by absolute altmetric counts (on the left side) and by altmetric coverage percentage (on the right side). Looking at left part of the table, it is observed that PHY and CHEM have 13 journals each in the list, followed by 12 journals from MED and 11 journals in BIO. The top few journals having highest altmetric absolute count are MUL discipline. Thus, in terms of absolute counts, MUL, MED, BIO are main disciplines. However, when the journals are sorted by altmetric coverage percentage, then journals from GEO and SS disciplines are also found listed. However, out of 50 journals in the list, 19 journals are still from MED discipline followed by BIO with 16 journals. Interestingly, ranking by altmetric coverage percentage results in only one journal from PHY figuring in the list. Thus, disciplinary variations are seen in this part of analysis as well.

To analyse the impact of journal even further, another sample data was extracted. This sample comprised of 50 journals, ranked by 2016 Impact Factor of journals. **Table 5** presents Web of Science article count for these top 50 journals along with their altmetric counts and coverage percentage. Here, most of the journals belong to either Medical Science or Biological Science. However, of particular interest would be the journals, which are from other disciplines. It is interesting to note that some of these journals have better coverage levels than the typical coverage level of that discipline. For example, the journal 'Nature Materials' has altmetric coverage percentage of 78.6%. Similarly, the journal 'Annual Review of Astronomy and Astrophysics' has an altmetric coverage percentage of 93.6%. Other journals like 'Nature Nanotechnology' has altmetric coverage percentage of 90.1% and 'Reviews of Modern Physics' has altmetric coverage percentage of 88.1%. This is higher coverage percentage than the overall coverage percentage of the respective disciplines. But at the same time several other journals like 'Progress in Materials Science', 'Progress in Polymer Science', 'Accounts of Chemical Research' and 'Behavioural and Brains Sciences' have altmetric coverage percentage around or below 50%. However, most of the journals in MED, BIO etc. continue to have higher altmetric coverage percentage. Therefore, it can be observed that there is a definite impact of the discipline of an article in its altmetric coverage. Articles from some disciplines have higher altmetric coverage. There are also some exceptions to this, where some journals in disciplines having relatively low altmetric coverage percentage, have higher altmetric coverage. Therefore, the journal has also some role to play in altmetric attention potential of an article.

**Table 4. Top 50 Journals Based on Altmetric Counts and Coverage Percentage**

| ***Sorted on Altmetric absolute Count*** | | | ***Sorted on Altmetric Coverage Percentage*** | | |
|---|---|---|---|---|---|
| *Journal* | *Discipline* | *TP_ALT* | *Journal* | *Discipline* | *Coverage %* |
| PLoS ONE | MUL | 15310 | PLoS Pathogens | BIO | 98.9 |
| Scientific Reports | MUL | 11017 | Nature | MUL | 94.3 |
| Nature Communications | MUL | 2849 | Atmospheric Chemistry & Physics | GEO | 93.6 |
| Proceedings of the National Academy of Sciences of the United States of America | MUL | 2821 | Bioinformatics | MED, BIO | 93.3 |
| British Medical Journal | MED | 2574 | Cell Reports | BIO | 93 |
| Oncotarget | BIO, MED | 2454 | American Journal of Public Health | MED | 93 |
| Angewandte Chemie. International Edition | CHE | 2374 | Angewandte Chemie. International Edition | CHE | 92.5 |
| Applied Physics Letters | PHY | 1873 | NeuroImage | MED | 92.2 |
| Dalton Transactions: An International Journal of Inorganic Chemistry | CHE | 1809 | Dalton Transactions: An International Journal of Inorganic Chemistry | CHE | 92.1 |
| RSC Advances | CHE | 1590 | Journal of Clinical Oncology | MED, BIO | 92 |
| Journal of the American Chemical Society | CHE | 1526 | Geophysical Research Letters | GEO | 90.9 |
| Physical Review B | PHY | 1489 | Nature Communications | MUL | 90.7 |
| Medicine | MED | 1454 | Blood | MED | 90.7 |
| Journal of Biological Chemistry | BIO | 1433 | Current Biology | BIO | 89.7 |
| Frontiers in Microbiology | BIO | 1419 | New England Journal of Medicine | MED | 89.6 |
| Frontiers in Plant Science | BIO | 1393 | Inorganic Chemistry | CHE | 89.4 |
| Physical Review D | PHY | 1373 | Proceedings of the National Academy of Sciences of the United States of America | MUL | 89 |
| Physical Review Letters | PHY | 1343 | British Medical Journal | MED | 88.9 |
| International Journal of Molecular Sciences | PHY, CHE, BIO | 1321 | Journal of Alzheimer's Disease | MED | 88.2 |
| Frontiers in Psychology | MED | 1286 | Clinical Cancer Research | MED | 87.8 |
| Chemistry - A European Journal | CHE | 1248 | Nucleic Acids Research | BIO, CHE | 87.2 |
| ACS Applied Materials & Interfaces | CHE, PHY | 1155 | PeerJ | BIO, MED | 86.4 |
| Monthly Notices of the Royal Astronomical Society | PHY | 1146 | Industrial & Engineering Chemistry Research | CHE, ENG | 86.3 |
| Geophysical Research Letters | GEO, | 1146 | BMC Genomics | BIO | 86.3 |
| The Astrophysical Journal | PHY | 1135 | Journal of Neuroscience | MED | 86.2 |
| Science | MUL | 1132 | Neurology | MED | 86.2 |
| Inorganic Chemistry | CHE | 1114 | PLoS Neglected Tropical Diseases | BIO | 85.8 |
| Chemical Communications | CHE | 1107 | Science | MUL | 85.7 |
| International Journal of Cardiology | MED | 1092 | JAMA: Journal of the American Medical Association | MED | 84.5 |
| Tumor Biology | MED,BIO | 1024 | eLife | BIO | 84.4 |
| Industrial & Engineering Chemistry Research | CHE, ENG | 992 | Pediatrics | MED | 84.4 |
| PeerJ | BIO, MED | 982 | Journal of Immunology | MED | 84.3 |
| Physical Chemistry Chemical Physics (PCCP) | CHE | 956 | Antimicrobial Agents and Chemotherapy | BIO | 83.1 |
| Journal of Physical Chemistry - Part C | CHE | 910 | Nutrients | AGR | 82.3 |
| BMJ Open | MED | 903 | Psychiatry Research | SS | 82.3 |
| Science of the Total Environment | ENV | 883 | Journal of Affective Disorders | SS | 81.7 |
| Blood | MED | 871 | Frontiers in Plant Science | BIO | 80.8 |
| Sensors (14248220) | PHY | 862 | Frontiers in Microbiology | BIO | 79.7 |
| Nature | MUL | 859 | Applied & Environmental Microbiology | BIO | 79.4 |
| Cell Reports | BIO | 857 | Journal of Dairy Science | AGR | 78.9 |
| Astronomy and Astrophysics | PHY | 839 | PLoS ONE | MUL | 77 |
| Physical Review A | PHY | 838 | Journal of Biological Chemistry | BIO | 76.4 |
| eLife | BIO | 830 | Water Research | ENV | 75.6 |
| Molecules | CHE | 813 | International Journal of Molecular Sciences | PHY, CHE, BIO | 75.3 |
| Journal of Neuroscience | MED | 810 | Journal of Medicinal Chemistry | MED, CHE | 74 |
| Journal of Clinical Oncology | MED, BIO | 807 | Journal of Virology | MED | 73.8 |
| NeuroImage | MED | 807 | Surgical Endoscopy | MED | 72.5 |
| Environmental Science & Technology | ENV | 806 | Journal of the American Chemical Society | CHE | 72.4 |
| Physical Review E | PHY | 804 | Nano Letters | MAR, CHE | 72.2 |
| Biochemical & Biophysical Research Communications | BIO, PHY | 783 | BMC Infectious Diseases | MED | 71.9 |

## Conclusion

This paper presents a comprehensive analytical study to explore whether there are apparent disciplinary variations in altmetric coverage of articles. A large sample of data from Web of Science along with corresponding data from altmetric.com is obtained and analysed. Results obtained show interesting patterns. Medical Sciences and Biology account for more than 50% of all instances in Altmetrics. In terms of coverage, disciplines like Biology, Medical Science and Multidisciplinary Sciences have more than 60% of their articles covered in Altmetrics, whereas disciplines like Engineering, Mathematics and Material Science have less than 25% of their articles covered in Altmetrics. The coverage percentages further vary across different altmetric platforms, with Twitter and Mendeley having much higher overall coverage than Facebook and News. Disciplinary variations in coverage are also found in different altmetric platforms, with variations as large as 7.5% for Engineering discipline to 55.7% for Multidisciplinary in Twitter. Some journals are also found to have a higher altmetric coverage in comparison to the average altmetric coverage level of that discipline, which shows that the source of publication may also have some impact on altmetric coverage of article.

**Table 5: Top 50 Journals (Sorted by 2016 Impact Factor (IF)) with corresponding WOS & Altmetric Values**

| *Journal* | *2016 IF* | *Discipline* | *TP_WOS* | *TP_ALT* | *Coverage %* |
|---|---|---|---|---|---|
| CA-A Cancer Journal For Clinicians | 131.723 | MED | 41 | 27 | 65.9 |
| New England Journal Of Medicine | 59.558 | MED | 838 | 751 | 89.6 |
| Nature Reviews Drug Discovery | 47.12 | MED, BIO | 162 | 127 | 78.4 |
| LANCET | 44.002 | MED | 522 | 460 | 88.1 |
| Nature Biotechnology | 43.113 | BIO | 173 | 126 | 72.8 |
| Nature Reviews Immunology | 39.416 | MED | 126 | 100 | 79.4 |
| Nature Materials | 38.891 | MAR | 238 | 187 | 78.6 |
| Nature Reviews Molecular Cell Biology | 38.602 | BIO | 126 | 122 | 96.8 |
| Nature | 38.138 | MUL | 911 | 859 | 94.3 |
| Annual Review of Astronomy and Astrophysics | 37.846 | PHY | 760 | 711 | 93.6 |
| JAMA-Journal of The American Medical Association | 37.684 | MED | 683 | 577 | 84.5 |
| Chemical Reviews | 37.369 | CHE | 260 | 179 | 68.8 |
| Nature Reviews Genetics | 35.898 | MED | 120 | 113 | 94.2 |
| Annual Review of Immunology | 35.543 | MED | 23 | 21 | 91.3 |
| Nature Nanotechnology | 35.267 | MAR | 141 | 127 | 90.1 |
| Science | 34.661 | MUL | 1321 | 1132 | 85.7 |
| Nature Reviews Cancer | 34.244 | MED | 87 | 85 | 97.7 |
| Chemical Society Reviews | 34.09 | CHE | 267 | 163 | 61 |
| Reviews Of Modern Physics | 33.177 | PHY | 42 | 37 | 88.1 |
| Living Reviews in Relativity | 32 | PHY | 2 | 1 | 50 |
| Nature Genetics | 31.616 | MED | 214 | 190 | 88.8 |
| Nature Photonics | 31.167 | PHY | 126 | 109 | 86.5 |
| Progress In Materials Science | 31.083 | MAR | 37 | 15 | 40.5 |
| Physiological Reviews | 30.924 | MED, SS | 39 | 30 | 76.9 |
| Nature Medicine | 30.357 | MED | 166 | 159 | 95.8 |
| Nature Reviews Neuroscience | 29.298 | MED | 125 | 107 | 85.6 |
| Cell | 28.71 | BIO | 537 | 505 | 94 |
| Nature Chemistry | 27.893 | CHE | 162 | 156 | 96.3 |
| Progress In Polymer Science | 27.184 | MAR | 35 | 13 | 37.1 |
| LANCET Oncology | 26.509 | MED | 324 | 239 | 73.8 |
| Energy & Environmental Science | 25.427 | ENV | 314 | 159 | 50.6 |
| Nature Methods | 25.328 | CHE | 182 | 159 | 87.4 |
| Nature Reviews Microbiology | 24.727 | BIO | 107 | 103 | 96.3 |
| Materials Science & Engineering R-Reports | 24.652 | MAR, PHY | 10 | 4 | 40 |
| Immunity | 24.082 | MED | 189 | 160 | 84.7 |
| Annual Review of Pathology-Mechanisms of Disease | 23.758 | BIO, MED | 23 | 19 | 82.6 |
| LANCET Neurology | 23.468 | MED | 108 | 78 | 72.2 |
| Cancer Cell | 23.214 | BIO, MED | 199 | 164 | 82.4 |
| Cell Stem Cell | 22.387 | BIO | 147 | 138 | 93.9 |
| Annual Review of Plant Biology | 22.131 | AGR, ENV, BIO | 67 | 1 | 1.5 |
| Accounts Of Chemical Research | 22.003 | CHE | 273 | 160 | 58.6 |
| Annual Review of Biochemistry | 21.407 | BIO | 28 | 3 | 10.7 |
| LANCET Infectious Diseases | 21.372 | MED | 246 | 213 | 86.6 |
| Journal Of Clinical Oncology | 20.982 | MED, BIO | 877 | 807 | 92 |
| Behavioral And Brain Sciences | 20.415 | MED | 262 | 141 | 53.8 |
| World Psychiatry | 20.205 | MED, SS | 91 | 81 | 89 |
| Cancer Discovery | 19.783 | MED | 126 | 104 | 82.5 |
| BMJ-British Medical Journal | 19.697 | MED | 2896 | 2574 | 88.9 |
| Nature Immunology | 19.381 | MED | 158 | 138 | 87.3 |
| Living Reviews in Solar Physics | 19.333 | PHY | 4 | 4 | 100 |

There are however some limitations of this study, which can be addressed in future work. The most important of them is the fact that disciplinary tagging of articles is based on ‘WC’ field of Web of Science, which classifies an article into a discipline based on its source of publication and not the actual article contents. It would, therefore, be interesting to take some large data sample, classify that into different disciplines using some Machine Learning Classifier (that processes article contents to tag it into a discipline), and then see if the disciplinary variation patterns are similar to those observed in this work. This would also establish the usefulness of Web of Science publication-source based disciplinary classification. Another interesting thing to explore could be to look in detail at the data from some particular journals (that have higher altmetric coverage) and to identify if there are some specific characteristics that helps a journal in attaining higher altmetric coverage, than the typical altmetric coverage level of that discipline.

**Acknowledgments**

The authors would like to acknowledge the access provided to data of altmetric.com by Stacy Konkiel. The authors also acknowledge the enabling support provided by the Indo-German Joint Research Project titled ‘Design of a Sciento-text Computational Framework for Retrieval and Contextual Recommendations of High-Quality Scholarly Articles’ (Grant No. DST/INT/FRG/DAAD/P-28/2017) for this work.